\documentclass[aps,pre,superscriptaddress,unsortedaddress,twocolumn,showpacs]{revtex4}
\usepackage{amsfonts}

\usepackage{graphicx}
\usepackage{epstopdf}
\usepackage{amsmath}  
\usepackage{amssymb}
\usepackage{verbatim}
\usepackage[ulem=normalem]{changes}
\usepackage{booktabs}
\usepackage{natbib}

\usepackage{xcolor}
\usepackage{graphicx}
\usepackage{tabulary}
\usepackage{tabularx}
\usepackage{threeparttable} 
\usepackage{array}
\newcolumntype{D}{ >{\centering\arraybackslash}m{7cm} }

\newcommand{\fpt}{first-passage time }

\newcommand{\co}{\color{orange}}
\newcommand{\cb}{\color{black}}

\begin{document}

\title{ Stochastic resetting in backtrack recovery by RNA polymerases}

\author{\'Edgar Rold\'an\footnote{Correspondence should be addressed to \'Edgar Rold\'an (Email: \href{mailto:edgar@pks.mpg.de}{edgar@pks.mpg.de}), and Stephan W. Grill (Email: \href{mailto:stephan.grill@biotec.tu-dresden.de}{stephan.grill@biotec.tu-dresden.de}).}}
\affiliation{Max Planck Institute for the Physics of Complex Systems, N{\"o}thnitzer Str. 38, 01187 Dresden, Germany.}
\affiliation{Center for Advancing Electronics Dresden, cfaed, Dresden, Germany.}
\affiliation{GISC - Grupo Interdisciplinar de Sistemas Complejos, Madrid, Spain.}

\author{Ana Lisica}
\affiliation{BIOTEC, Technische Universit{\"a}t Dresden, Tatzberg 47/49, 01307 Dresden, Germany.}
\affiliation{Max Planck Institute of Molecular Cell Biology and Genetics, Pfotenhauerstra{\ss}e 108, 01307 Dresden, Germany.}
\author{Daniel S\'anchez-Taltavull}
\affiliation{Regenerative Medicine Program, Ottawa Hospital Research Institute, Ottawa, K1H 8L6, Canada.}
\author{Stephan W. Grill$^{*}$}
\affiliation{Max Planck Institute for the Physics of Complex Systems, N{\"o}thnitzer Str. 38, 01187 Dresden, Germany.}
\affiliation{BIOTEC, Technische Universit{\"a}t Dresden, Tatzberg 47/49, 01307 Dresden, Germany.}
\affiliation{Max Planck Institute of Molecular Cell Biology and Genetics, Pfotenhauerstra{\ss}e 108, 01307 Dresden, Germany.}

\medskip

\begin{abstract}
 Transcription is a key process in gene expression, in which RNA polymerases produce a complementary RNA copy from a DNA template. RNA polymerization is frequently interrupted by backtracking, a process in which polymerases perform a  random walk  along the DNA template.  Recovery of polymerases from the  transcriptionally-inactive backtracked state is determined by a kinetic competition between 1D diffusion and RNA cleavage. Here we  describe backtrack recovery as a continuous-time random walk, where the time for a polymerase to recover from a backtrack of a given depth  is described as a  first-passage time of a random walker to reach an absorbing state. We represent RNA cleavage as a stochastic resetting process, and derive exact expressions for the recovery time distributions and mean recovery times from a given initial backtrack depth for both continuous and discrete-lattice descriptions of the random walk. We show that recovery time statistics do not depend on the discreteness of the DNA lattice when the rate of 1D diffusion is large compared to the rate of cleavage. 
\end{abstract}

\maketitle

\section{Introduction}

\begin{figure}[h]
\centering
\includegraphics[width= 8cm]{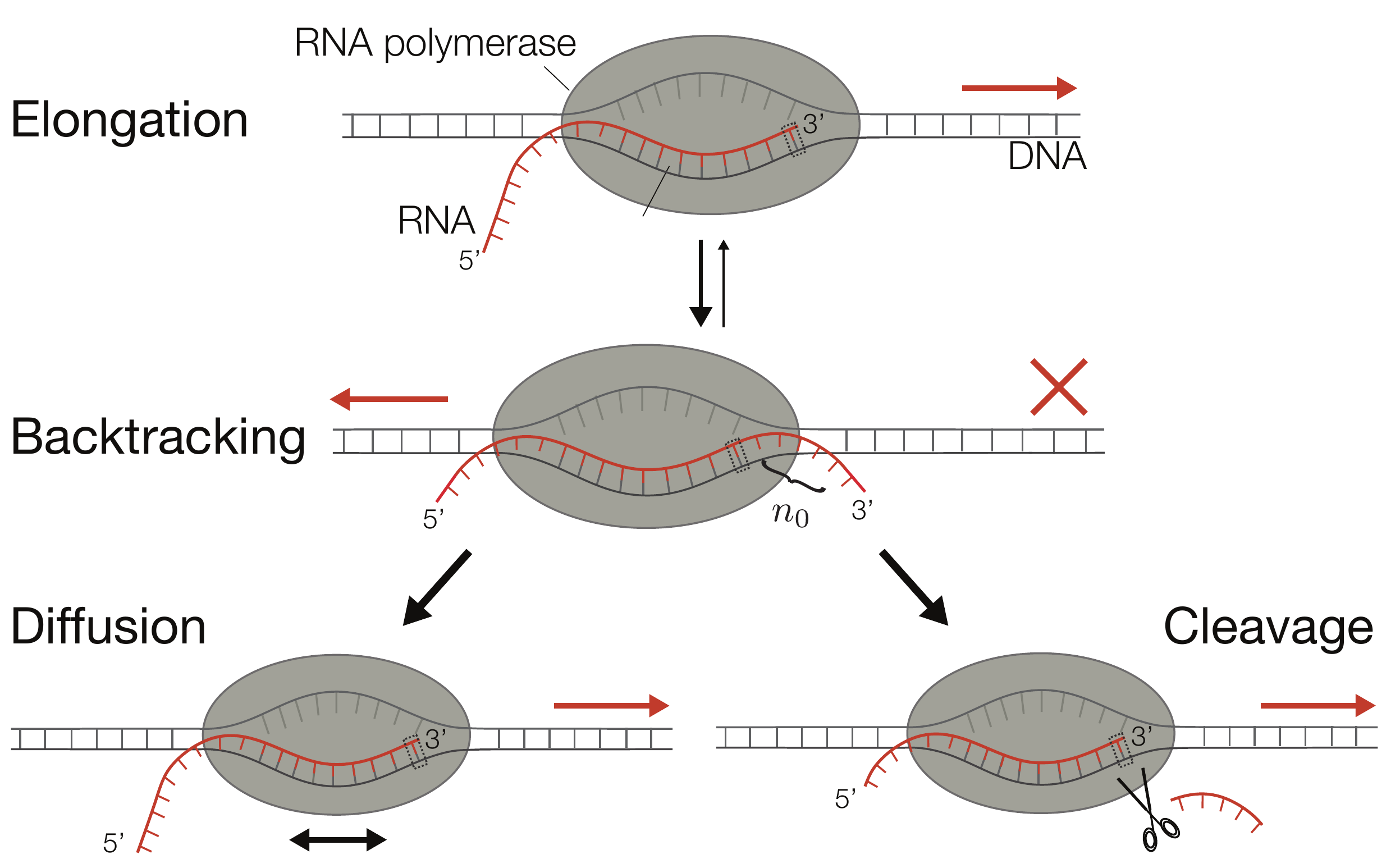}

\caption[Scheme of transcription elongation, backtracking and backtrack recovery processes.] {\small\label{fig:fig1} {\bfseries Scheme of transcription elongation (top), backtracking (middle) and backtrack recovery (bottom) processes.}  
The 3'-end of the RNA is aligned with the active site of the polymerase (dashed square) during elongation, but displaced in the backtracked state. This makes the polymerase transcriptionally inactive, as it can not add new nucleotides to the 3'- end of the RNA. Here, $n_0$ represents the number of backtracked nucleotides. Backtrack recovery can proceed through 1D diffusion of the polymerase along the DNA template, until the 3Õ-end of the RNA realigns with the active site or through cleaving of the backtracked RNA and production of a new 3'-end that is aligned with the active site.}
\end{figure}

Transcription of genetic information from DNA into RNA is the first step of gene expression and is fundamental for cellular regulation. The process is performed by macromolecular enzymes called RNA polymerases that move stepwise along a DNA template and produce a complementary RNA (see Fig.~\ref{fig:fig1}). Transcription elongation is often interrupted by pausing and backtracking, a reverse movement of the RNA polymerase on the DNA template that displaces the RNA 3' end from the active site and leaves the enzyme transcriptionally inactive \cite{Nudler:1997uz,Komissarova:1997vs,Kettenberger:2003iq,Wang:2009kaa,Cheung:2011gg}. The polymerase recovers from a backtrack when it realigns the 3' end of the RNA with its active site. 

In a backtrack, polymerases perform random walk on the DNA template~\cite{Galburt:2007bf}. The recovery of the polymerase from a backtracked state, i.e. {\em backtrack recovery},  results from the kinetic competition between two mechanisms~\cite{lisica2016mechanisms}:   polymerases can either recover by performing  a random walk  along the DNA until it returns to the elongation competent state \cite{Shaevitz:2003wd,Galburt:2007bf,Depken:2009wa,Hodges:2009cl,Dangkulwanich:2013hi,Ishibashi:2014cd} or by cleavage of the backtracked RNA which generates a new RNA 3' end in the active site \cite{Kuhn:2007en,Walmacq:2009cr,Chedin:1998iq}. The cleavage reaction can be performed by intrinsic cleavage mechanisms or it can be assisted by a transcription factor, TFIIS \cite{izban:2013vq,Fish:2002ug,Ruan:2011hl}.

 The stochastic motion of the RNA polymerase in a backtrack was previously measured with single-molecule optical tweezers and described using continuous time Markov processes ~\cite{Shaevitz:2003wd,Galburt:2007bf,Hodges:2009cl,Anonymous:2012ds,Dangkulwanich:2013hi,Ishibashi:2014cd}. Specifically, backtracking has been treated as a diffusion process in continuous space \cite{Galburt:2007bf} but also as a hopping process over a discrete lattice of nucleotides  \cite{Depken:2009wa,Hodges:2009cl,Klopper:2010fy, Dangkulwanich:2013hi,sahoo2013backtracking,Ishibashi:2014cd,Schweikhard:2014bx}. However, it remains unclear which aspects of the backtracking process depend on the discreteness of the position lattice and which can be described with a diffusion process.

Here we present both discrete and continuous-space descriptions of backtrack recovery and investigate to which extent a diffusion process is a good approximation of the polymerase dynamics during a backtrack. We present a solvable stochastic model of RNA polymerase backtrack recovery that includes both diffusion and cleavage and study its main statistical features. The process shares similarities with recent development on diffusion processes with {\em stochastic resetting} introduced in Ref.~\cite{evans2011diffusion}. In such problems a particle undergoes Brownian diffusion but can also stochastically reset its position~\cite{evans2011diffusion,majumdarres,gupta2014fluctuating,durang2014statistical,nagar2015diffusion,pal2015diffusion,rotbart2015michaelis,reuveni2015optimal}. The mean \fpt to an absorber can be determined analytically, which depends on the statistics of resetting~\cite{evans2011diffusion,majumdarres,nagar2015diffusion,pal2015diffusion}. A backtracked RNA polymerase undergoes a random walk to the elongation competent state while also resetting its position via cleavage, and we here determine the \fpt properties of this variant of a 'diffusion with stochastic resetting' process.

In experiments, deep backtracks are readily identified, and it is possible to determine accurately the time it takes the polymerase to recover from a backtrack of a certain depth~\cite{lisica2016mechanisms}.  Therefore, we here determine the recovery time $\tau_{\rm rec}$, defined as the \fpt of a random walker to reach an absorbing barrier with the walker starting at a given 'initial' backtracking depth. We derive exact expressions of relevant statistics, such as the mean time to recover from a backtrack, or mean recovery time, for both continuous and discrete stochastic models. Remarkably, we find that for the case when the hopping rate is much larger than the cleavage rate  both discrete and continuous  descriptions can be used concurrently to describe the statistics of backtrack recovery from short and long initial backtrack depths.


\begin{figure}
\centering
\includegraphics[width= .45 \textwidth]{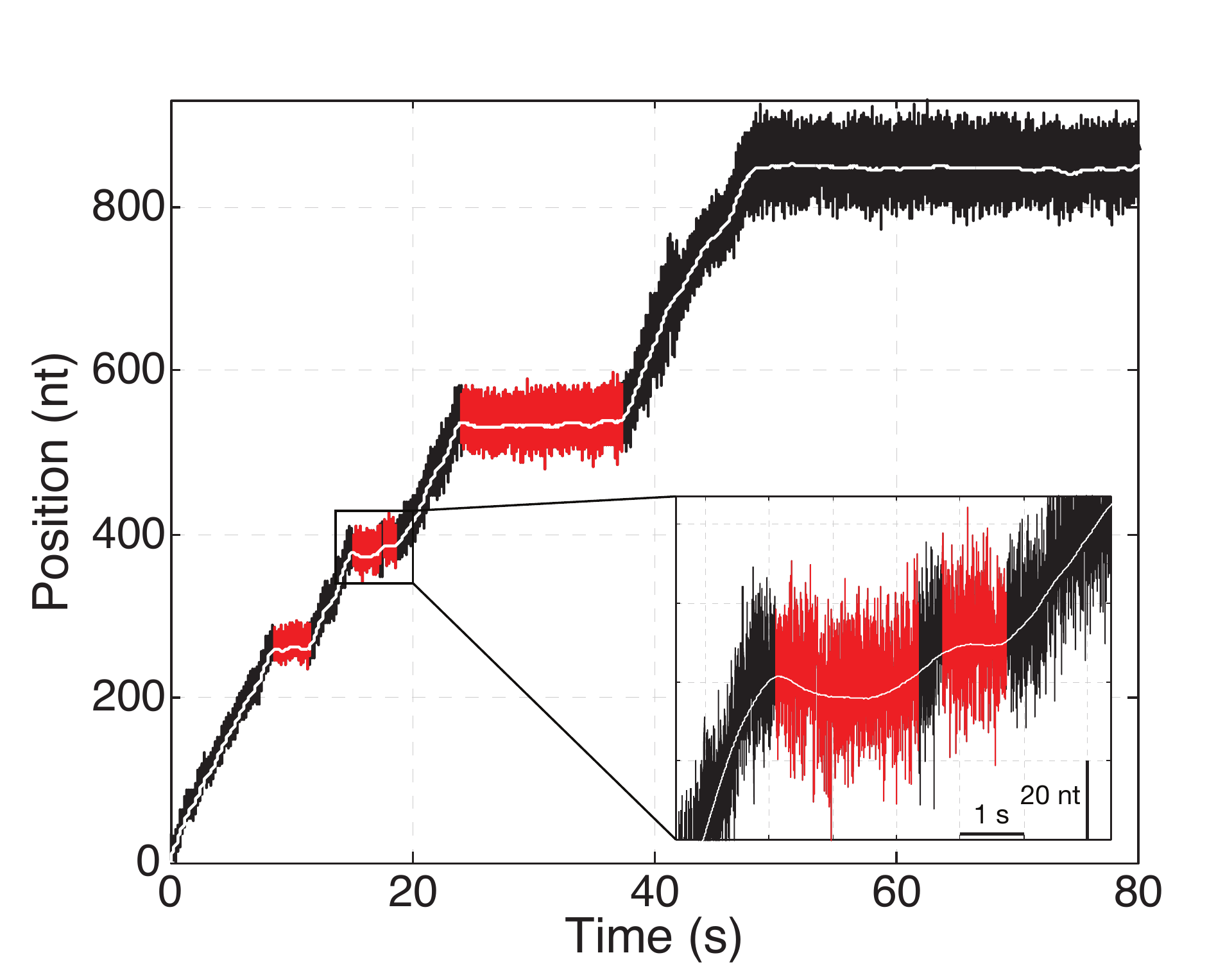}
\caption[Experimental traces of transcription elongation by RNA polymerase I.] {\small\label{fig:trace} {\bfseries Experimental traces of transcription elongation by RNA polymerase I.} 
Position of the polymerase on a DNA template (in nucleotides, nt) as a function of time obtained in a single-molecule experiment. The regions highlighted in red correspond to pauses and the inset shows a zoomed view of one of the pauses, which includes a backtrack. The original data (black) is obtained with the experimental setup described in~\cite{Jahnel:2011uv,lisica2016mechanisms}, using a sampling rate of $1$ kHz.}
\end{figure}

\section{Discrete model: hopping process with cleavage}\cb

\begin{figure}
\centering
\includegraphics[width= .45 \textwidth]{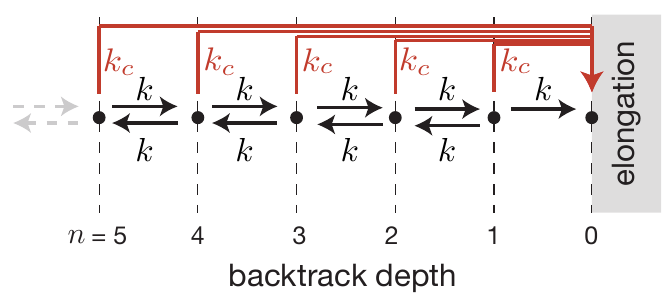}
\caption[Stochastic model of backtrack recovery: hopping process with cleavage] {\small\label{fig:modelBrown} {\bfseries Stochastic model of backtrack recovery: hopping process with cleavage.} 
Each state $n$ represents the number of backtracked nucleotides. The stochastic motion of the polymerases in a backtrack is described as a continuous-time hopping process between adjacent states with hopping rate $k$. Cleavage is represented as a stochastic reset to the elongation state with rate $k_{c}$. The recovery time from an initial backtrack depth $n_0$ is given by the first-passage time to the absorbing elongation state, $n=0$.}
\end{figure}

We first describe  the recovery of an RNA polymerase from a backtrack as a continuous-time 1D hopping process on a semi-infinite discrete lattice. Each state of the lattice $n\in [1,2,3,\cdots,\infty)$  represents the number of nucleotides backtracked by the polymerase  (see Fig.~\ref{fig:modelBrown}). For example, $n(t)=3$ means that at time $t$  the polymerase has backtracked $3$ nucleotides. In our model, polymerases can jump between adjacent states with hopping rate $k$ and can cleave an arbitrarily long RNA transcript with a cleavage rate $k_c$. 
We consider that no external forces  bias the hopping rates of the polymerase on the lattice. Cleavage is represented by an instantaneous jump or stochastic reset~\cite{evans2011diffusion,majumdarres,gupta2014fluctuating,durang2014statistical,nagar2015diffusion,pal2015diffusion,rotbart2015michaelis,reuveni2015optimal} to the elongation-competent state located in $n=0$. The elongation-competent state is considered as an absorbing state because the probability to backtrack after cleavage is very low~\cite{Dangkulwanich:2013hi}. Our discrete model is a variant of the hopping models introduced by Depken {\em et al.} in Refs.~\cite{Depken:2009wa,Depken:2013dj}.

The time evolution of the position of the polymerase can be described in a Master equation formalism~\cite{gardiner2009}. The probability of the polymerase to be at state $n$ at time $t\geq 0$ is given by $p_n(t)$. We consider the initial condition $p_n(0)=\delta_{n,n_0}$, that is, polymerases are initially positioned at $n_0\geq 1$. The dynamics of the probability of the polymerase to be at a given state at time $t$ is described by the following Master equation:
\begin{eqnarray}
\frac{\text{d}{p}_1(t)}{\text{d}t} &=& k\, p_2 (t) - (2k + k_c)\, p_1 (t)\quad , \label{eq:me1} \\
\frac{\text{d}{p}_n(t)}{\text{d}t}  &=& k\, p_{n+1} (t) - (2k + k_c )\, p_n (t)+ k\, p_{n-1} (t)\; , \label{eq:me3}
\end{eqnarray}
where $n\geq 2$. The elongation state $n=0$ is an absorber.  Recent experiments showed that the elongation rate from $n=0$ is more than 10 times faster than the rate of backtracking by one nucleotide~\cite{Dangkulwanich:2013hi}, hence we neglect the possibility to make a jump from $n=0$ to $n=1$. Hence, the recovery time is the \fpt of the polymerase to reach the absorber located in $n=0$. 

Equations~(\ref{eq:me1}-\ref{eq:me3}) can be solved exactly (see Appendix A). Using the exact solution of the model we now derive exact expressions for the recovery time distribution and the mean recovery time of a polymerase from a given initial backtracked state.


\subsection{Recovery time distribution}
Next, we derive an analytical expression for the recovery time distribution from an initial backtrack depth, $n_0$. We introduce a generating function
\begin{equation}
G (t,z) \equiv \sum_{n=1}^{\infty} p_n (t) z^{n-1}\quad .
\end{equation}
For $z=0$, the generating function gives the probability to be in $n=1$ at time $t$, $G(t,0) = p_1 (t)$. For $z=1$, the generating function equals the survival probability $S(t;n_0)$ at time $t$ starting from $n_0$,  $G(t,1)=\sum_{n=1}^{\infty} p_n (t)=S(t;n_0)$.

Using the generating function, the full set of Master equations [Eqs.~(\ref{eq:me1}-\ref{eq:me3})] can be rewritten as a single ordinary differential equation for the generating function
\begin{equation}
\frac{\partial G (t,z)}{\partial t} = \left[kz - (2k+k_c) +\frac{k}{z}\right] G (t,z)- \frac{k}{z}G (t,0)\quad .
\label{eq:genfunode}
\end{equation}
The initial condition  $p_{n} (0) = \delta_{n,n_0}$  can be  expressed in terms of the generating function as $G (0,z) = \sum_{n=1}^{\infty} p_n(0) z^{n-1} = \sum_{n=1}^{\infty} \delta_{n,n_0} z^{n-1} = z^{n_0-1}$.
The solution of Eq.~\eqref{eq:genfunode} with this initial condition is
\begin{eqnarray}
\lefteqn{G (t,z;n_0) }&&\nonumber\\
&= & \exp\left[ \left(kz - (2k+k_c) +\frac{k}{z}\right)t\right]  \nonumber\\
& \times &\left[ z^{n_0-1} - \frac{k}{z}\int_0^t    e^{-(kz - (2k+k_c) + k/z)s}      G (s,0) \text{d}s  \right] .\label{eq:GFsolution}
\end{eqnarray}

We next define $\Phi(\tau_{\rm rec};n_0)\, \text{d}\tau_{\rm rec}$ as the probability of  a polymerase to recover from an initial backtracked position $n_0$ in the time interval $[\tau_{\rm rec},\tau_{\rm rec}+\text{d}\tau_{\rm rec}]$. To calculate $\Phi(\tau_{\rm rec};n_0)$, we use the fact that a polymerase can exit a backtrack by hopping (from state $n=1$ with rate $k$) or by cleavage (from any state with rate $k_c$). The probability density of the polymerase to reach the absorbing state at time $\tau_{\rm rec}$ is then given by 
\begin{equation}
\Phi (\tau_{\rm rec};n_0) = k\, G (\tau_{\rm rec},0;n_0) + k_c\, G(\tau_{\rm rec},1;n_0)\quad .
\label{eq:phit}
\end{equation}

The probability to be at the state $1$ in $\tau_{\rm rec}$, $G (\tau_{\rm rec},0;n_0)$, equals to (see Appendix A)
\begin{equation}
G(\tau_{\rm rec},0;n_0) = e^{-(2k+k_c)\tau_{\rm rec}} \frac{n_0\,I_{n_0}(2k\tau_{\rm rec})}{k\tau_{\rm rec}}\quad ,
\label{eq:p0}
\end{equation}
where $I_{n_0}$ is the $n_0 -$th order modified Bessel function of the first kind~\cite{abramowitz1964handbook}. 

The survival probability in $\tau_{\rm rec}$, $S(\tau_{\rm rec};n_0)=G(\tau_{\rm rec},1;n_0)$ is given by 
\begin{equation}
G(\tau_{\rm rec},1;n_0) = e^{-k_c \tau_{\rm rec}} \left[ 1 - k \int_0^{\tau_{\rm rec}} e^{-2ks} \frac{n_0\,I_{n_0}(2ks)}{ks} \text{d}s\right]\,,
\end{equation}
which yields
\begin{equation}
G(\tau_{\rm rec},1;n_0)  = e^{-k_c \tau_{\rm rec}} \left[ 1-\frac{(k \tau_{\rm rec})^{n_0}}{n_0 \Gamma (n_0)}H(\tau_{\rm rec};n_0)\right]\quad,
\label{eq:eta1}
\end{equation}
where $\Gamma$ is the Gamma function and $H$ equals to
\begin{eqnarray}
\lefteqn{H(\tau_{\rm rec};n_0)}&&\\
&=& \, _2F_2\left[\left\{n_0,n_0+\frac{1}{2}\right\};\left\{n_0+1,2 n_0+1\right\};-4 k \tau_{\rm rec} \right]\quad,\nonumber
\end{eqnarray}
where $_2F_2$ is a generalized hypergeometric function (see Ref.~\cite{abramowitz1964handbook}). \cb
 The recovery time distribution is obtained by substituting~\eqref{eq:p0} and~\eqref{eq:eta1} in~\eqref{eq:phit}, and given by
\begin{eqnarray}
\lefteqn{\Phi(\tau_{\rm rec};n_0)} &&\nonumber \\
&=& e^{-(2k+k_c)\tau_{\rm rec}} \frac{n_0\,I_{n_0}(2k\tau_{\rm rec})}{\tau_{\rm rec}}\nonumber \\
&+& k_c e^{-k_c \tau_{\rm rec}} \left[ 1- \frac{(k \tau_{\rm rec})^{n_0} \; H(\tau_{\rm rec};n_0)}{n_0 \Gamma (n_0)}\right]\quad .\label{eq:etd}
\end{eqnarray}
\begin{figure*}
\centering
\includegraphics[width= 0.8\textwidth]{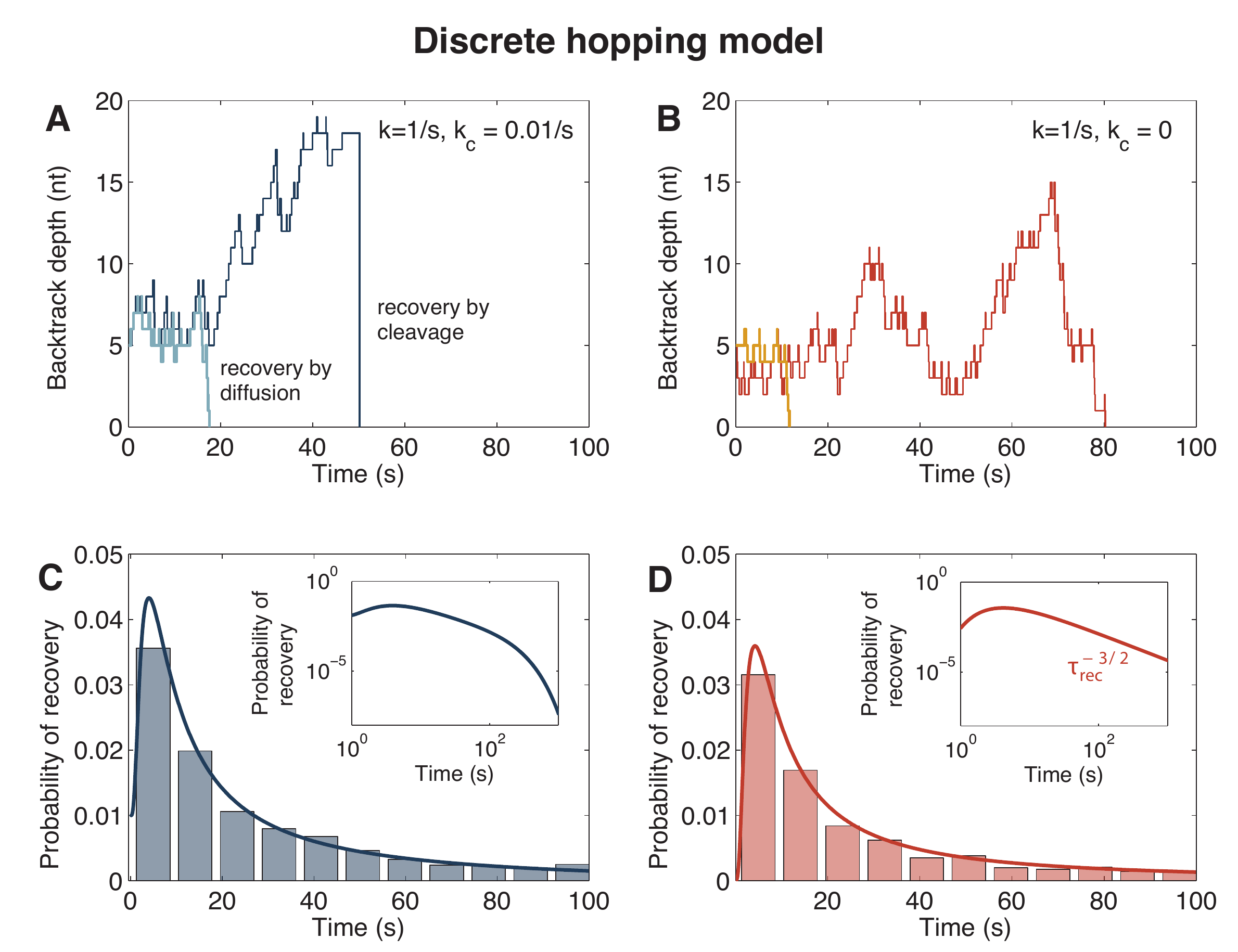}
\caption[Stochastic trajectories of the discrete hopping model and recovery time distributions.] {\small\label{fig:simulations} {\bfseries Stochastic trajectories of the discrete hopping model and recovery time distributions.} \textbf{A)} Sample trajectories of the hopping model with diffusion and cleavage ($k=1/s$, $k_c=0.01/s$) simulated using the Gillespie algorithm. The light blue trajectory represents a polymerase that recovers by diffusion, and the dark blue trajectory a polymerase that recovers by cleavage. \textbf{B)} Sample trajectories for the discrete model with only diffusion, $k=1/s$, $k_c=0$, obtained using the Gillespie algorithm. \textbf{C)} Recovery time probability density for the case where $k=1/s$ and $k_c=0.01/s$. The bars are obtained from histograms of $1000$ numerical simulations and the curve is the exact expression given by Eq.~\eqref{eq:etd}. The inset shows a log-log plot of the recovery time distribution for long recovery times. \textbf{D)} Numerical and analytical probability density of the recovery time for the case where, $k=1/s$, $k_c=0$. The inset shows the tail $\tau_{\rm rec}^{-3/2}$ of the distribution at long times. In all cases the initial backtrack depth was set to $n_0=5$. }
\end{figure*}
In the absence of cleavage, $k_c=0$, the recovery time distribution becomes
\begin{equation}
\Phi(\tau_{\rm rec};n_0) = e^{-2kt} \frac{n_0\,I_{n_0}(2kt)}{t}\quad.
\label{eq:etdiff}
\end{equation}
Note that if the polymerase is initially at $n_0=1$, we obtain the probability density for a pause of duration $\tau_{\rm rec}$,  
\begin{equation}
\Psi(\tau_{\rm rec}) = e^{-2k\tau_{\rm rec}} \frac{I_1(2k\tau_{\rm rec})}{\tau_{\rm rec}}\quad ,
\end{equation}
in agreement with Depken {\em et al.}~\cite{Depken:2009wa}.


To verify our model, we perform numerical simulations of the hopping process with cleavage using Gillespie algorithm~\cite{gillespie1976general} (Fig. \ref{fig:simulations}A, B). From our simulations, we calculate \fpt distributions to the elongation state and compare it with the recovery time distribution derived in Eq.~\eqref{eq:etd} (Fig. \ref{fig:simulations}C, D). In the presence of cleavage, recovery can happen from arbitrarily deep backtracks. Cleavage prevents backtracks of large duration as shown by the sharp cutoff of the \fpt distribution at large times (Fig. \ref{fig:simulations}C, inset). In the absence of cleavage, deep backtracks are recovered at very large times, with a power-law tail $\Phi(\tau_{\rm rec};n_0)\sim \tau_{\rm rec}^{-3/2}$ (Fig. \ref{fig:simulations}D, inset).  The  probability density function obtained from numerical simulations in both cleavage assisted (Fig.  \ref{fig:simulations}C) and cleavage deficient case (Fig. \ref{fig:simulations}D) agree with the theoretical expression of the recovery time distribution derived here in Eq.~\eqref{eq:etd}.

\subsection{Mean recovery time}
\label{sec:discmean}

The mean recovery time $\langle\tau_{\rm rec}\rangle$ is a useful statistic that can be measured experimentally in single-molecule experiments. Moreover, the mean recovery time can provide a quantitative measure of kinetic rates of backtrack recovery, as shown in Ref.~\cite{lisica2016mechanisms}. The mean recovery time can be obtained from Eq.~\eqref{eq:etd}, and equals to
\begin{equation}
\langle\tau_{\rm rec}\rangle= \frac{1}{k_c}\left[  1 - \left(  \frac{\sqrt{\frac{4k}{k_c} +1}-1}{\sqrt{\frac{4k}{k_c} +1}+1}   \right)^{n_0}    \right]\quad.
\label{eq:tauexit}
\end{equation}
 We introduce the following characteristic scales of time and backtrack position,
\begin{eqnarray}
n_c &=&  \sqrt{\frac{4k}{k_c}}\label{eq:snd}\quad , \\
\tau_c &=& \frac{1}{k_c}\label{eq:std}\quad .
\end{eqnarray}
The mean recovery time then simplifies to
\begin{equation}
\langle\tau_{\rm rec}\rangle= \tau_c \left[  1 - \left(  \frac{\sqrt{n_c^2+1}-1}{\sqrt{n_c^2 +1}+1}   \right)^{n_0}    \right]\quad .
\label{eq:MRTD}
\end{equation}

For initial positions that are not large compared to $n_c$, i.e. when $n_0\leq n_c$, the mean recovery time given by Eq.~\eqref{eq:MRTD} depends linearly on the initial backtrack depth
\begin{equation}
\langle\tau_{\rm rec}\rangle= \tau_c \ln \left( \frac{\sqrt{n_c^2+1}+1}{\sqrt{n_c^2 +1}-1} \right) n_0 + O(n_0^2)\quad.
\label{eq:meandiscshort}
\end{equation}
When the initial position is much larger than the characteristic backtrack position $n_c$, i.e. when $n_0\gg n_c$, the mean recovery  time saturates to $\langle\tau_{\rm rec}\rangle\to \tau_c$.  

If the hopping and cleavage rates are equal ($k=k_c$), the mean recovery time given by Eq.~\eqref{eq:MRTD}  can be rewritten in terms of the Golden ratio $\varphi=(\sqrt{5}+1)/2$ and  the Golden ratio conjugate $\Phi = (\sqrt{5}-1)/2$
\begin{eqnarray}
\langle\tau_{\rm rec}\rangle_{k=k_c} = \tau_c \left[  1 -  \left(  \frac{\Phi}{\varphi}   \right)^{n_0}    \right]\quad .
\end{eqnarray}
The   duration of a transcriptional pause, or equivalently the mean recovery time from $n_0=1$ is, for the case where $k=k_c$ equal to $\langle\tau_{\rm rec}\rangle_{k=k_c,n=1} = \tau_c / \varphi$.
 
\begin{figure}
\centering
\includegraphics[width= 7cm]{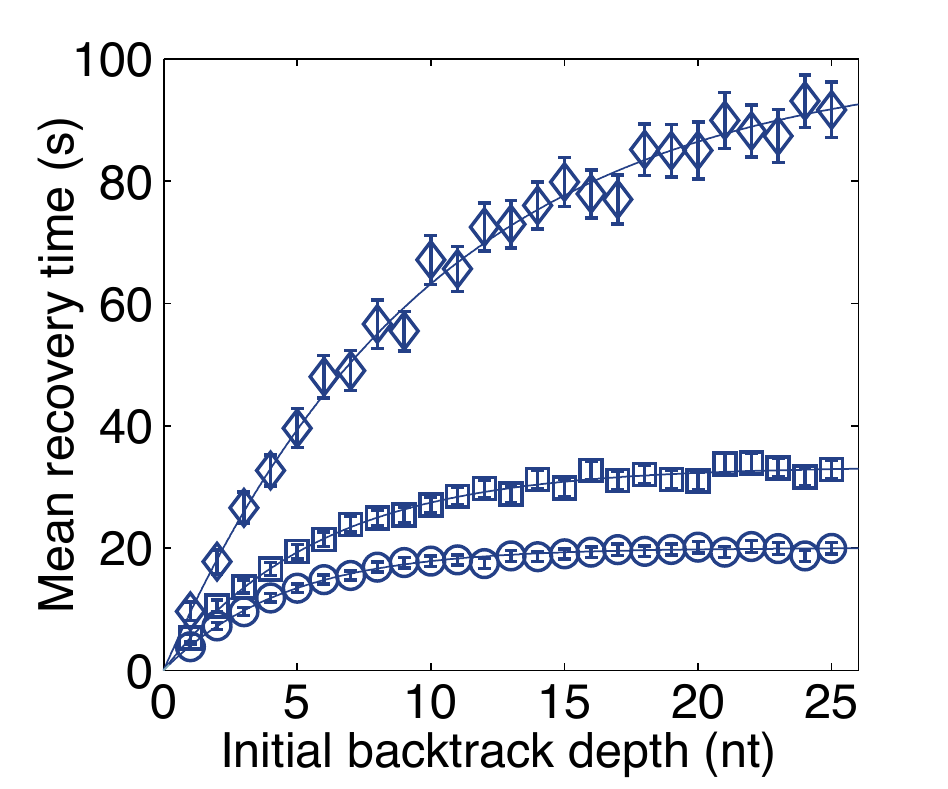}
\caption[Mean recovery time as a function of the backtrack depth, discrete model.] {\small\label{fig:fig4} {\bfseries Mean recovery time as a function of the backtrack depth, discrete model.} Recovery time averaged over $1000$ numerical simulations (symbols) of recovery from different initial backtrack depths. Diffusion rate was set to $k=1/\rm s$ in all cases and $k_c = 0.01/\rm s$ (diamonds), $k_c = 0.03/\rm s$ (squares) and $k_c = 0.05/\rm s$ (circles). Error bars are standard errors of the mean with $90\%$ statistical significance. The solid curves are obtained with the analytical expression given by Eq.~\eqref{eq:tauexit} for the specific values of $n_0$, $k$ and $k_c$ in this example.}
\end{figure}

Figure~\ref{fig:fig4} shows   the analytical expression of the mean recovery time in the discrete model~\eqref{eq:tauexit} compared to the average recovery time obtained from numerical simulations. The mean recovery time increases with increasing initial backtrack depth and saturates at $1/k_c$ for deep initial backtracks. The saturation of the mean recovery time at large $n_0$  was observed experimentally for Pol II~recovery assisted with TFIIS~\cite{lisica2016mechanisms}.

In the absence of cleavage, the mean recovery time is not bounded, yielding $\langle\tau_{\rm rec}\rangle=\infty$. Alternative statistics should  therefore be considered to characterize the recovery in the absence of cleavage, such as the mode or the median recovery times.  

\section{Continuous model: diffusion process with cleavage}\cb

To address which features of the backtrack recovery process depend on the details of the 1D lattice of the DNA template, we now consider a continuous-space model where the motion of the polymerase is described by a diffusion process with a stochastic resetting~\cite{evans2011diffusion} to the elongation state due to RNA cleavage. 
 Such model can be envisioned as the continuous limit of the model in Fig.~\ref{fig:modelBrown}.

We consider that the position of the polymerase, $x$, is a continuous random variable. We define $\rho(x,t|x_0,0)dx$ as the probability of a polymerase to be in the interval $[x,x+dx]$ at time $t$, given that it was at $x_0$ at time $0$. In this continuous-space description the probability density $\rho(x,t|x_0,0)$ evolves in time according to a Fokker-Planck equation with a diffusion term and a sink term,
\begin{equation}
\frac{\partial\rho(x,t|x_0,0)}{\partial t} = D \frac{\partial^2\rho(x,t|x_0,0)}{\partial x^2} - k_c \rho(x,t|x_0,0)\quad,
\label{eq:FPE}
\end{equation}
where we assume $x>0$.  Equation~\eqref{eq:FPE} results from taking the continuous limit in Eq.~\eqref{eq:me3} and defining $x=an$, $x_0=an_0$ and the diffusion coefficient $D=a^2 k$, with $a=0.34 \,\rm nm$ the distance between two nucleotides.  The solution of the Fokker-Planck equation~\eqref{eq:FPE} with initial condition $\rho(x,0|x_0,0)=\delta(x-x_0)$ and the absorbing boundary condition $\rho(0,t|x_0,0)=0$  for $x>0$ is given by~\cite{redner2001guide}: 
\begin{equation}
\rho(x,t|x_0,0) = \frac{e^{-k_c t}}{\sqrt{4\pi D t}}\left[  e^{-(x-x_0)^2/4Dt} -e^{-(x+x_0)^2/4Dt}   \right]\quad.
\label{eq:fpesol}
\end{equation} 

The recovery time probability density is given by the probability density flux to $x=0$ due to diffusion, plus the probability flux due to cleavage,
\begin{eqnarray}
\lefteqn{\Phi (\tau_{\rm rec};x_0) } &&\nonumber\\
&=& \Phi_{\rm diff} (\tau_{\rm rec};x_0) +\Phi_{\rm c} (\tau_{\rm rec};x_0)\\
&=&D \left . \frac{\partial \rho(x,\tau_{\rm rec}|x_0,0)}{\partial x}\right|_{x=0} + k_{c} S(\tau_{\rm rec};x_0)\quad,
\end{eqnarray}
where $\Phi_{\rm diff} (\tau_{\rm rec};x_0)\text{d}\tau_{\rm rec}$ is the probability to recover by diffusion in the time interval $[\tau_{\rm rec},\tau_{\rm rec}+\text{d}\tau_{\rm rec}]$ and $\Phi_{\rm c} (\tau_{\rm rec};x_0)\text{d}\tau_{\rm rec}$ is the probability to recover by cleavage in the time interval $[\tau_{\rm rec},\tau_{\rm rec}+\text{d}\tau_{\rm rec}]$. The probability density flux  across $x=0$ due to diffusion equals to
\begin{eqnarray}
\lefteqn{\Phi_{\rm diff} (\tau_{\rm rec};x_0) } && \nonumber\\
&=& D \left . \frac{\partial \rho(x,\tau_{\rm rec}|x_0,0)}{\partial x}\right|_{x=0}\label{eq:diffflux} \\
& =& e^{-k_c \tau_{\rm rec}}\frac{x_0}{\sqrt{4\pi D \tau_{\rm rec}^3}} e^{-x_0^2/4D\tau_{\rm rec}}\quad,
\label{eq:phidiff}
\end{eqnarray}
where the spatial derivative in Eq.~\eqref{eq:diffflux} is a derivative from the right. The survival probability at time $\tau_{\rm rec}$ can be calculated by integrating the probability of the polymerase to be at time $\tau_{\rm rec}$ in $x>0$,
\begin{eqnarray}
\lefteqn{S(\tau_{\rm rec};x_0) } &&\nonumber\\
&=& \int_0^\infty \rho(x,\tau_{\rm rec}|x_0,0)\, \text{d}x \\
& =& e^{-k_c \tau_{\rm rec}} \text{erf}\left( \frac{x_0}{\sqrt{4D\tau_{\rm rec}}}  \right) \quad,
\label{eq:survivalprob}
\end{eqnarray}
where erf is the error function. 
The probability density $R(\tau_{\rm rec};x_0)$ of recovery from an initial backtrack depth $x_0$ in a time $\tau_{\rm rec}$, (referred to as the recovery probability) is then given by
\begin{equation}
R(\tau_{\rm rec};x_0)= 1 - S(\tau_{\rm rec};x_0) = 1 - e^{-k_c \tau_{\rm rec}} \text{erf}\left( \frac{x_0}{\sqrt{4D\,\tau_{\rm rec}}} \right).
\label{eq:probrecWT}
\end{equation}
For the case $k_c=0$, the recovery probability simplifies to
\begin{equation}
R(\tau_{\rm rec};x_0)=  \text{erfc}\left( \frac{x_0}{\sqrt{4D\,\tau_{\rm rec}}} \right)\quad,
\label{eq:probrecMUT}
\end{equation}
where erfc is the complementary error function. From Eq.~\eqref{eq:survivalprob}, we obtain the probability density flux through $x=0$ via cleavage:
\begin{equation}
\Phi_{\rm c} (\tau_{\rm rec};x_0)= k_c S(\tau_{\rm rec};x_0) = k_c e^{-k_c \tau_{\rm rec}} \text{erf}\left( \frac{x_0}{\sqrt{4D\tau_{\rm rec}}}  \right)\quad.
\label{eq:phic}
\end{equation}

We obtain an exact expression for the recovery time distribution in the continuous-space model by adding Eq.~\eqref{eq:phidiff} to~\eqref{eq:phic}
\begin{eqnarray}
\lefteqn{\Phi(\tau_{\rm rec};x_0)} &&\nonumber\\
&=&  e^{-k_c \tau_{\rm rec}}\frac{x_0}{\sqrt{4\pi D \tau_{\rm rec}^3}} e^{-x_0^2/4D\tau_{\rm rec}} \nonumber\\
&+&  k_c e^{-k_c \tau_{\rm rec}} \text{erf}\left( \frac{x_0}{\sqrt{4D\tau_{\rm rec}}}  \right)\quad.
\label{eq:phiSI}
\end{eqnarray}
We now write Eq.~\eqref{eq:phiSI} scaling time with respect to $\tau_c=1/k_c$ and the initial position with respect to $x_c = \sqrt{4D/k_c}$ similarly to Eqs.~\eqref{eq:snd} and~\eqref{eq:std} in the hopping model. In units of a scaled time $\mathsf{t}_{\rm rec}= \tau_{\rm rec}/\tau_c$ and a scaled initial position $\mathsf{x}_0 = x_0/x_c$, we obtain a universal expression
\begin{equation}
\Phi(\mathsf{t}_{\rm rec};\mathsf{x}_0)=e^{-\mathsf{t}_{\rm rec}} \frac{\mathsf{x}_0}{\sqrt{\pi \mathsf{t}_{\rm rec}^3}} e^{-\mathsf{x}_0^2/\mathsf{t}_{\rm rec}} + e^{-\mathsf{t}_{\rm rec}} \text{erf}\left( \frac{\mathsf{x}_0}{\sqrt{\mathsf{t}_{\rm rec}}}  \right)\quad.
\label{eq:scaledRTd}
\end{equation}

\begin{figure*}
\centering
\includegraphics[width= 0.8\textwidth]{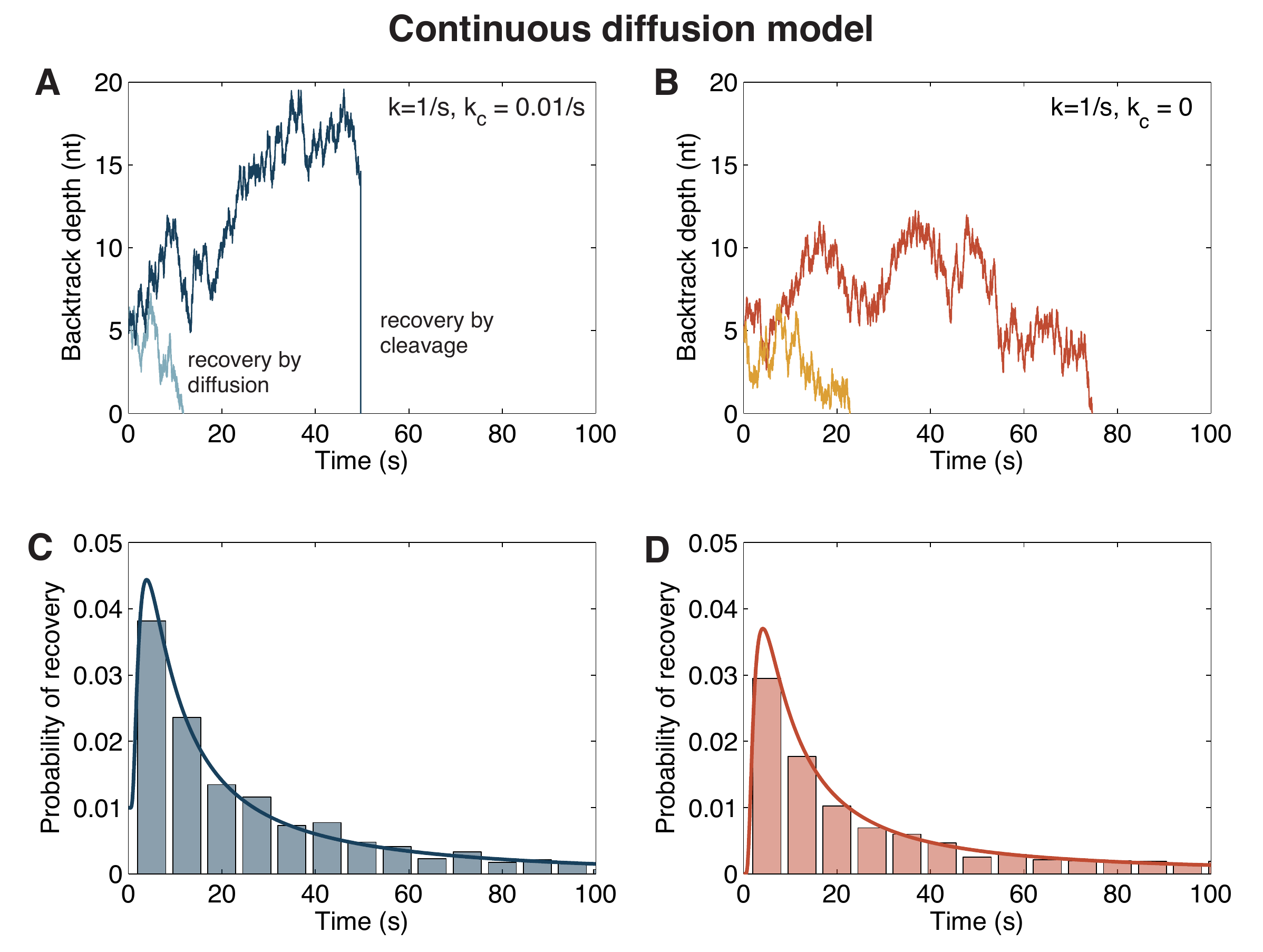}
\caption[Stochastic trajectories of the continuous diffusion model and recovery time distributions.] {\small\label{fig:fig2} {\bfseries Stochastic trajectories of the continuous diffusion model and recovery time distributions.} \textbf{A)} Sample trajectories of the continuous-space model with diffusion and cleavage ($k=1/s$, $k_c=0.01/s$) simulated using Euler numerical scheme for a discrete-time Langevin equation. The light blue trajectory represents a polymerase that recovers by diffusion, and the dark blue trajectory a polymerase that recovers by cleavage. \textbf{B)} Sample trajectories for the continuous model with only diffusion, $k=1/s$, $k_c=0$ obtained with the same numerical integration scheme. \textbf{C)}  Recovery time probability density for the case where $k=1/s$ and $k_c=0.01/s$. The bars are histograms obtained from $1000$ numerical simulations and the curve is the exact expression given by Eq.~\eqref{eq:phiSI}. \textbf{D)} Recovery time probability density for the case where $k=1/s$ and $k_c=0$. The bars are histograms obtained from $1000$ numerical simulations and the curve is the exact expression given by Eq.~\eqref{eq:phiSI} with $k_c=0$. In all simulations, the simulation time step was set to $\Delta t=1\,\rm ms$ and the initial distance to $x_0 = 5\,\rm nt $.
\label{fig:simulationscont}
}
\end{figure*}

To test the validity of the analytical expression for the recovery time distribution~\eqref{eq:scaledRTd}, we perform numerical simulations of the continuous model (see Fig.~\ref{fig:simulationscont}). The following overdamped Langevin equation describes the evolution of the backtracked distance at time $t$ in continuous space, denoted as $x(t)$, $\text{d}x(t)/\text{d}t = \xi (t)$ where $\xi(t)$ models a stochastic force that drives the polymerase forward or backward. The stochastic force is described by a delta-correlated Gaussian white noise with zero mean $\langle \xi(t)\rangle=0$ and an amplitude proportional to the diffusion coefficient, $\langle \xi(t)\xi(t')\rangle = 2D \delta (t-t')$. Cleavage events are modelled as a stochastic resetting process~\cite{evans2011diffusion} whose probability to occur in a time $t$ is exponential $P_{\rm cleav}(t)=k_c e^{-k_c t}$. We implement numerical simulations of  the Langevin equation using an Euler discrete-time numerical integration scheme with $\Delta t =1\,\rm ms$, which is one order of magnitude smaller 
than any characteristic time of backtrack recovery  given by the inverse of cleavage or diffusion rates \cite{lisica2016mechanisms}. The results shown in Fig.~\ref{fig:simulationscont} validate the exact expression obtained for the recovery time distribution given by Eq.~\eqref{eq:phiSI}  both in the presence and in the absence of cleavage. The recovery time distributions obtained for the same initial backtrack distance $x_0=5$ have the same shape as those obtained in the discrete-space description [cf. Fig.~\ref{fig:simulations}].


\subsection{Mean recovery time}
 In the continuous model,  the mean recovery time can be  obtained by calculating the mean value of the first-passage distribution [Eq.~\eqref{eq:phiSI}]
\begin{equation}
\langle\tau_{\rm rec}\rangle= \frac{1}{k_c} \left[   1-  e^{- x_0/\sqrt{D/k_c}}   \right]\quad,
\label{eq:meanRTSI}
\end{equation}
or equivalently
\begin{equation}
\langle\tau_{\rm rec}\rangle= \tau_c \left[   1-  e^{ - 2 x_0/x_c }      \right]\quad.
\label{eq:MRTC}
\end{equation}
Note that, the mean recovery time can be also calculated  by  a different route, using the backward Fokker-Planck equation  together with  the Laplace transform of the survival probability (see Appendix B).

Equations~\eqref{eq:meanRTSI} and~\eqref{eq:MRTC} show that the mean recovery time for  deep  initial backtracks $(x_0\gg x_c)$ \cb saturates to $\tau_c$. Our results indicate that recovery happens mostly by diffusion for shallow backtracks, where $x_0\ll x_c$, and mostly by cleavage for deep backtracks, where $x_0\gg x_c$. 
For shallow initial backtracks  $(x_0\leq x_c)$,  the mean recovery time scales linearly with $x_0$,
\begin{equation}
\frac{\langle\tau_{\rm rec}\rangle}{\tau_c} = \frac{x_0}{x_c/2} + O(x_0^2)\quad,
\label{eq:meanRTcontshort}
\end{equation}
 similarly to the  mean recovery time in the discrete hopping model  [see Eq. \eqref{eq:meandiscshort}].

Taking the limit  $n_c\gg 1$ in the expression for the mean recovery time in the discrete model [Eq.~\eqref{eq:MRTD}], we obtain
\begin{equation}
\langle\tau_{\rm rec}\rangle\simeq \tau_c\left[1-  e^{-2n_0/(\sqrt{n_c^2+1}+1)}     \right] \simeq \tau_c\left[1-  e^{-2x_0/x_c}     \right]\quad,\label{eq:TA}
\end{equation}
where we have used $x_0=an_0$ and $x_c=a n_c$.  Note that this expression is equal to the mean recovery time in the continuous model [Eq.~\eqref{eq:MRTC}]. Hence, the mean recovery times in the  discrete and continuous description coincide for large characteristic depth. 

 A comparison between the mean recovery time in the discrete and continuous model illustrates that both agree well for $n_c\geq1$. In this regime, the polymerase typically performs a large number of jumps prior to recovery. Hence, the lattice does not impact on the mean recovery time, even for shallow initial backtracks. 
 
Notably, the limit $n_c\geq 1$ is in agreement with the experimental data obtained from single molecule experiments with RNA~polymerases~II, where $k\sim 1~\text{s}^{-1}$ and $k_c \sim (0.01-0.1)~\text{s}^{-1}$ \cite{Hodges:2009cl,Zamft:2012es,Dangkulwanich:2013hi,Ishibashi:2014cd,lisica2016mechanisms}, yielding $n_c>1$. Therefore, for reported values of diffusion and cleavage rates, the diffusion approximation can be used without loss of generality, with the advantage of providing a simpler mathematical framework with respect to the hopping process. \co

\cb

%
%


\begin{figure}
\centering
\includegraphics[width= 6.5cm]{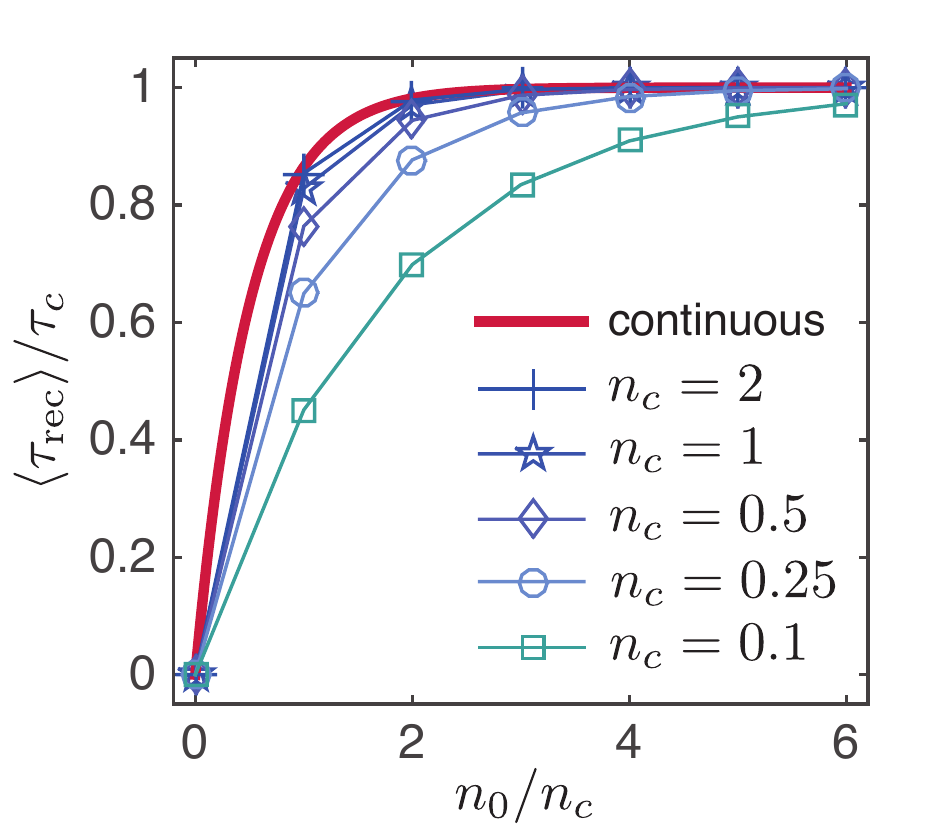}
\caption[ Scaled mean recovery time as a function of the scaled initial backtrack depth for discrete and continuous models.] {\small\label{fig:fig6} {\bfseries Scaled mean recovery time as a function of the scaled initial backtrack depth for discrete and continuous models.} Scaled mean recovery time $\langle\tau_{\rm rec}\rangle/\tau_c$ as a function of the scaled initial backtrack depth $n_0/n_c$ for the discrete-space hopping model [Eq.~\eqref{eq:MRTD}, blue symbols] and as a function of $x_0/x_c$ for the continuous-space diffusion model [Eq.~\eqref{eq:MRTC}, magenta curve]. 
\label{fig:fig6}}
\end{figure}

\cb
\section{Discussion}

We have provided exact results on the statistics of the time needed for an RNA polymerase to recover from an arbitrary initial backtrack depth using  discrete (hopping) and continuous-space (diffusion) stochastic descriptions. We have presented a road-map for the calculation of the \fpt distribution for a continuous-time random walk with an absorbing state, which models RNA polymerase backtrack recovery with high fidelity. Both hopping and diffusion models provide similar recovery time distributions, with the majority of differences in the short recovery times and a complete overlap for long recovery times (see Table I for a summary of the main results).

 We show that  both discrete and continuous description can be used concurrently for backtrack recovery analysis for short and long backtracks when the characteristic distance $n_c = 2\sqrt{k/k_c}$ is greater than one.  This corresponds to cases where the hopping rate $k$ is larger than the cleavage rate $k_{c}$ and  is in good agreement with estimated rates of RNA polymerase  backtracking \cite{Dangkulwanich:2013hi,Ishibashi:2014cd,lisica2016mechanisms}.  Future work in the framework of stochastic resetting will  have  to be done to consider the case where polymerases can only cleave  until  a critical  backtrack  distance, as recently found in single-molecule experiments~\cite{lisica2016mechanisms}.  Single-molecule optical tweezers transcription experiments of RNA~polymerase backtracking \cite{Shaevitz:2003wd,Abbondanzieri:2005dz,Galburt:2007bf,Hodges:2009cl,Dangkulwanich:2013hi,Ishibashi:2014cd,Schweikhard:2014bx,lisica2016mechanisms} would allow to experimentally validate the stochastic models provided here and quantify the backtrack diffusion and cleavage rates of these enzymes. \\





\section{acknowledgments}

We are thankful to Shamik Gupta, Izaak Neri, Federico Vazquez,  Masatoshi Nishikawa, Cl\'elia de Mulatier and Francesc Font-Clos  for helpful discussions.  ER acknowledges financial support from Spanish Government, grants ENFASIS (FIS2011-22644) and TerMic (FIS2014-52486-R). ER and DST are thankful to Centro de Ciencias de Benasque Pedro Pascual (Benasque, Spain) for its hospitality. S.W.G. was supported by the EMBO Young Investigator Program, the Paul Ehrlich Foundation and grant no.~281903 from the European Research Council.

\newpage
\vspace{5cm}

\begin{widetext}
\renewcommand{\arraystretch}{1.75}
\begin{table}[!htp]
\begin{threeparttable}
\caption{Summary of expressions for the probability distribution of the recovery time and  the mean recovery time   from a given initial backtrack depth in the hopping model with cleavage $\Phi(\tau_{\rm rec};n_0)$ and in the diffusion model with cleavage $\Phi(\tau_{\rm rec};x_0)$ with initial backtrack depths $n_0$ and $x_0$ respectively. Here, $k$ is the hopping rate, $D$ is the diffusion coefficient  and $H(\tau_{\rm rec};n_0)= \, _2F_2(\{n_0,n_0+1/2\};\{n_0+1,2 n_0+1\};-4 k \tau_{\rm rec} )$.} {\small\label{table:table1}}\centering 
\small
\begin{tabularx}{\textwidth}{ |X|D|D| }

\multicolumn{3}{c}{ Discrete hopping model}\\\hline
  & Diffusion and cleavage ($k>0 \; ; \;k_c>0$) & Only diffusion ($k>0 \; ; \;k_c=0$) \\\hline
$\Phi(\tau_{\rm rec};n_0)$ &  $e^{-(2k+k_c)\tau_{\rm rec}} \frac{n_0\,I_{n_0}(2k\tau_{\rm rec})}{\tau_{\rm rec}} + k_c e^{-k_c \tau_{\rm rec}} \left[ 1- \frac{(k \tau_{\rm rec})^{n_0} \; H(\tau_{\rm rec};n_0)}{n_0 \Gamma (n_0)}\right]$ & $e^{-(2k+k_c)\tau_{\rm rec}} \frac{n_0\,I_{n_0}(2k\tau_{\rm rec})}{\tau_{\rm rec}}$ \\\hline
$\langle\tau_{\rm rec}\rangle$  &  $ \frac{1}{k_c} \left[  1 - \left(  \frac{\sqrt{(4k/k_c)+1}-1}{\sqrt{(4k/k_c) +1}+1}   \right)^{n_0}    \right]$ &  $\infty$   \\\hline

\multicolumn{3}{c}{ Continuous diffusion model}  \\\hline
  & Diffusion and cleavage ($D>0 \; ; \;k_c>0$)  & Only diffusion ($D>0 \; ; \;k_c=0$)\\\hline
$\Phi(\tau_{\rm rec};x_0)$ &  $e^{-k_c \tau_{\rm rec}}\frac{x_0}{\sqrt{4\pi D \tau_{\rm rec}^3}} e^{-x_0^2/4D\tau_{\rm rec}} +  k_c e^{-k_c \tau_{\rm rec}} \text{erf}\left( \frac{x_0}{\sqrt{4D\tau_{\rm rec}}}  \right)$ & $\frac{x_0}{\sqrt{4\pi D \tau_{\rm rec}^3}} e^{-x_0^2/4D\tau_{\rm rec}} $ \\\hline
$\langle\tau_{\rm rec}\rangle$   & $\frac{1}{k_c} \left[   1-  e^{- x_0/\sqrt{D/k_c}}   \right]$	& $\infty$   \\\hline

  \hline
  \end{tabularx}
    \end{threeparttable}
\end{table} 
\end{widetext}

\bibliographystyle{naturemag}

\begin{thebibliography}{10}
\expandafter\ifx\csname url\endcsname\relax
  \def\url#1{\texttt{#1}}\fi
\expandafter\ifx\csname urlprefix\endcsname\relax\def\urlprefix{URL }\fi
\providecommand{\bibinfo}[2]{#2}
\providecommand{\eprint}[2][]{\url{#2}}

\bibitem{Nudler:1997uz}
\bibinfo{author}{Nudler, E.}, \bibinfo{author}{Mustaev, A.},
  \bibinfo{author}{Goldfarb, A.} \& \bibinfo{author}{Lukhtanov, E.}
\newblock \bibinfo{title}{The {RNA-DNA} hybrid maintains the register of
  transcription by preventing backtracking of {RNA} polymerase}.
\newblock \emph{\bibinfo{journal}{Cell}} \textbf{\bibinfo{volume}{89}},
  \bibinfo{pages}{33--41} (\bibinfo{year}{1997}).

\bibitem{Komissarova:1997vs}
\bibinfo{author}{Komissarova, N.} \& \bibinfo{author}{Kashlev, M.}
\newblock \bibinfo{title}{Transcriptional arrest: $\textit{E. coli}$ {RNA}
  polymerase translocates backward, leaving the 3'\--end of the {RNA} intact
  and extruded}.
\newblock \emph{\bibinfo{journal}{Proc. Natl. Acad. Sci. U.S.A.}}
  \textbf{\bibinfo{volume}{94}}, \bibinfo{pages}{1755--1760}
  (\bibinfo{year}{1997}).

\bibitem{Kettenberger:2003iq}
\bibinfo{author}{Kettenberger, H.}, \bibinfo{author}{Armache, K.-J.} \&
  \bibinfo{author}{Cramer, P.}
\newblock \bibinfo{title}{Architecture of the {RNA} polymerase~{II-TFIIS}
  complex and implications for m{RNA} cleavage}.
\newblock \emph{\bibinfo{journal}{Cell}} \textbf{\bibinfo{volume}{114}},
  \bibinfo{pages}{347--357} (\bibinfo{year}{2003}).

\bibitem{Wang:2009kaa}
\bibinfo{author}{Wang, D.} \emph{et~al.}
\newblock \bibinfo{title}{{Structural basis of transcription: backtracked RNA
  polymerase II at 3.4 angstrom resolution}}.
\newblock \emph{\bibinfo{journal}{Science}} \textbf{\bibinfo{volume}{324}},
  \bibinfo{pages}{1203--1206} (\bibinfo{year}{2009}).

\bibitem{Cheung:2011gg}
\bibinfo{author}{Cheung, A.~C.} \& \bibinfo{author}{Cramer, P.}
\newblock \bibinfo{title}{{Structural basis of RNA polymerase II backtracking,
  arrest and reactivation}}.
\newblock \emph{\bibinfo{journal}{Nature}} \textbf{\bibinfo{volume}{471}},
  \bibinfo{pages}{249--253} (\bibinfo{year}{2011}).

\bibitem{Galburt:2007bf}
\bibinfo{author}{Galburt, E.~A.} \emph{et~al.}
\newblock \bibinfo{title}{{Backtracking determines the force sensitivity of
  RNAP II in a factor-dependent manner}}.
\newblock \emph{\bibinfo{journal}{Nature}} \textbf{\bibinfo{volume}{446}},
  \bibinfo{pages}{820--823} (\bibinfo{year}{2007}).

\bibitem{lisica2016mechanisms}
\bibinfo{author}{Lisica, A.} \emph{et~al.}
\newblock \bibinfo{title}{Mechanisms of backtrack recovery by rna polymerases i
  and ii}.
\newblock \emph{\bibinfo{journal}{Proceedings of the National Academy of
  Sciences}} \textbf{\bibinfo{volume}{113}}, \bibinfo{pages}{2946--2951}
  (\bibinfo{year}{2016}).

\bibitem{Shaevitz:2003wd}
\bibinfo{author}{Shaevitz, J.~W.}, \bibinfo{author}{Abbondanzieri, E.~A.},
  \bibinfo{author}{Landick, R.} \& \bibinfo{author}{Block, S.~M.}
\newblock \bibinfo{title}{{Backtracking by single RNApolymerase molecules
  observedat near-base-pair resolution}}.
\newblock \emph{\bibinfo{journal}{Nature}} \textbf{\bibinfo{volume}{426}},
  \bibinfo{pages}{684--687} (\bibinfo{year}{2003}).

\bibitem{Depken:2009wa}
\bibinfo{author}{Depken, M.}, \bibinfo{author}{Galburt, E.~A.} \&
  \bibinfo{author}{Grill, S.~W.}
\newblock \bibinfo{title}{{The origin of short transcriptional pauses}}.
\newblock \emph{\bibinfo{journal}{Biophys. J.}} \textbf{\bibinfo{volume}{96}},
  \bibinfo{pages}{2189--2193} (\bibinfo{year}{2009}).

\bibitem{Hodges:2009cl}
\bibinfo{author}{Hodges, C.}, \bibinfo{author}{Bintu, L.},
  \bibinfo{author}{Lubkowska, L.}, \bibinfo{author}{Kashlev, M.} \&
  \bibinfo{author}{Bustamante, C.}
\newblock \bibinfo{title}{{Nucleosomal fluctuations govern the transcription
  dynamics of RNA polymerase II}}.
\newblock \emph{\bibinfo{journal}{Science}} \textbf{\bibinfo{volume}{325}},
  \bibinfo{pages}{626--628} (\bibinfo{year}{2009}).

\bibitem{Dangkulwanich:2013hi}
\bibinfo{author}{Dangkulwanich, M.} \emph{et~al.}
\newblock \bibinfo{title}{{Complete dissection of transcription elongation
  reveals slow translocation of RNA polymerase II in a linear ratchet
  mechanism}}.
\newblock \emph{\bibinfo{journal}{eLife}} \textbf{\bibinfo{volume}{2}}
  (\bibinfo{year}{2013}).

\bibitem{Ishibashi:2014cd}
\bibinfo{author}{Ishibashi, T.} \emph{et~al.}
\newblock \bibinfo{title}{{Transcription factors IIS and IIF enhance
  transcription efficiency by differentially modifying RNA polymerase pausing
  dynamics}}.
\newblock \emph{\bibinfo{journal}{Proc. Natl. Acad. Sci. U.S.A.}}
  \textbf{\bibinfo{volume}{111}}, \bibinfo{pages}{3419--3424}
  (\bibinfo{year}{2014}).

\bibitem{Kuhn:2007en}
\bibinfo{author}{Kuhn, C.-D.} \emph{et~al.}
\newblock \bibinfo{title}{{Functional architecture of RNA polymerase I}}.
\newblock \emph{\bibinfo{journal}{Cell}} \textbf{\bibinfo{volume}{131}},
  \bibinfo{pages}{1260--1272} (\bibinfo{year}{2007}).

\bibitem{Walmacq:2009cr}
\bibinfo{author}{Walmacq, C.} \emph{et~al.}
\newblock \bibinfo{title}{{Rpb9 subunit controls transcription fidelity by
  delaying NTP sequestration in RNA polymerase II}}.
\newblock \emph{\bibinfo{journal}{J. Biol. Chem.}}
  \textbf{\bibinfo{volume}{284}}, \bibinfo{pages}{19601--19612}
  (\bibinfo{year}{2009}).

\bibitem{Chedin:1998iq}
\bibinfo{author}{Ch{\'e}din, S.}, \bibinfo{author}{Riva, M.},
  \bibinfo{author}{Schultz, P.}, \bibinfo{author}{Sentenac, A.} \&
  \bibinfo{author}{Carles, C.}
\newblock \bibinfo{title}{{The RNA cleavage activity of RNA polymerase III is
  mediated by an essential TFIIS-like subunit and is important for
  transcription termination}}.
\newblock \emph{\bibinfo{journal}{Genes Dev.}} \textbf{\bibinfo{volume}{12}},
  \bibinfo{pages}{3857--3871} (\bibinfo{year}{1998}).

\bibitem{izban:2013vq}
\bibinfo{author}{Izban, M.~G.} \& \bibinfo{author}{Luse, D.~S.}
\newblock \bibinfo{title}{{The RNA polymerase II ternary complex cleaves the
  nascent transcript in a 3'--5' direction in the presence of elongation factor
  SII.}}
\newblock \emph{\bibinfo{journal}{Genes Dev.}} \textbf{\bibinfo{volume}{6}},
  \bibinfo{pages}{1342--1356} (\bibinfo{year}{1992}).

\bibitem{Fish:2002ug}
\bibinfo{author}{Fish, R.~N.} \& \bibinfo{author}{Kane, C.~M.}
\newblock \bibinfo{title}{{Promoting elongation with transcript cleavage
  stimulatory factors}}.
\newblock \emph{\bibinfo{journal}{Biochim. Biophys. Acta}}
  \textbf{\bibinfo{volume}{1577}}, \bibinfo{pages}{287--307}
  (\bibinfo{year}{2002}).

\bibitem{Ruan:2011hl}
\bibinfo{author}{Ruan, W.}, \bibinfo{author}{Lehmann, E.},
  \bibinfo{author}{Thomm, M.}, \bibinfo{author}{Kostrewa, D.} \&
  \bibinfo{author}{Cramer, P.}
\newblock \bibinfo{title}{{Evolution of two modes of intrinsic RNA polymerase
  transcript cleavage}}.
\newblock \emph{\bibinfo{journal}{J. Biol. Chem.}}
  \textbf{\bibinfo{volume}{286}}, \bibinfo{pages}{18701--18707}
  (\bibinfo{year}{2011}).

\bibitem{Anonymous:2012ds}
\bibinfo{author}{Larson, M.~H.} \emph{et~al.}
\newblock \bibinfo{title}{{Trigger loop dynamics mediate the balance between
  the transcriptional fidelity and speed of RNA polymerase II}}.
\newblock \emph{\bibinfo{journal}{Proc. Natl. Acad. Sci. U.S.A.}}
  \textbf{\bibinfo{volume}{109}}, \bibinfo{pages}{6555--6560}
  (\bibinfo{year}{2012}).

\bibitem{Klopper:2010fy}
\bibinfo{author}{Klopper, A.~V.}, \bibinfo{author}{Bois, J.~S.} \&
  \bibinfo{author}{Grill, S.~W.}
\newblock \bibinfo{title}{{Influence of secondary structure on recovery from
  pauses during early stages of RNA transcription}}.
\newblock \emph{\bibinfo{journal}{Physical Review E}}
  \textbf{\bibinfo{volume}{81}}, \bibinfo{pages}{1--4} (\bibinfo{year}{2010}).

\bibitem{sahoo2013backtracking}
\bibinfo{author}{Sahoo, M.} \& \bibinfo{author}{Klumpp, S.}
\newblock \bibinfo{title}{{Backtracking dynamics of RNA polymerase: pausing and
  error correction}}.
\newblock \emph{\bibinfo{journal}{Journal of Physics: Condensed Matter}}
  \textbf{\bibinfo{volume}{25}}, \bibinfo{pages}{374104}
  (\bibinfo{year}{2013}).

\bibitem{Schweikhard:2014bx}
\bibinfo{author}{Schweikhard, V.} \emph{et~al.}
\newblock \bibinfo{title}{{Transcription factors TFIIF and TFIIS promote
  transcript elongation by RNA polymerase II by synergistic and independent
  mechanisms}}.
\newblock \emph{\bibinfo{journal}{Proc. Natl. Acad. Sci. U.S.A.}}
  \textbf{\bibinfo{volume}{111}}, \bibinfo{pages}{6642--6647}
  (\bibinfo{year}{2014}).

\bibitem{evans2011diffusion}
\bibinfo{author}{Evans, M.~R.} \& \bibinfo{author}{Majumdar, S.~N.}
\newblock \bibinfo{title}{Diffusion with stochastic resetting}.
\newblock \emph{\bibinfo{journal}{Phys. Rev. Lett.}}
  \textbf{\bibinfo{volume}{106}}, \bibinfo{pages}{160601}
  (\bibinfo{year}{2011}).

\bibitem{majumdarres}
\bibinfo{author}{Evans, M.~R.} \& \bibinfo{author}{Majumdar, S.~N.}
\newblock \bibinfo{title}{Diffusion with optimal resetting}.
\newblock \emph{\bibinfo{journal}{Journal of Physics A: Mathematical and
  Theoretical}} \textbf{\bibinfo{volume}{44}}, \bibinfo{pages}{435001}
  (\bibinfo{year}{2011}).

\bibitem{gupta2014fluctuating}
\bibinfo{author}{Gupta, S.}, \bibinfo{author}{Majumdar, S.~N.} \&
  \bibinfo{author}{Schehr, G.}
\newblock \bibinfo{title}{Fluctuating interfaces subject to stochastic
  resetting}.
\newblock \emph{\bibinfo{journal}{Phys. Rev. Lett.}}
  \textbf{\bibinfo{volume}{112}}, \bibinfo{pages}{220601}
  (\bibinfo{year}{2014}).

\bibitem{durang2014statistical}
\bibinfo{author}{Durang, X.}, \bibinfo{author}{Henkel, M.} \&
  \bibinfo{author}{Park, H.}
\newblock \bibinfo{title}{The statistical mechanics of the
  coagulation--diffusion process with a stochastic reset}.
\newblock \emph{\bibinfo{journal}{Journal of Physics A: Mathematical and
  Theoretical}} \textbf{\bibinfo{volume}{47}}, \bibinfo{pages}{045002}
  (\bibinfo{year}{2014}).

\bibitem{nagar2015diffusion}
\bibinfo{author}{Nagar, A.} \& \bibinfo{author}{Gupta, S.}
\newblock \bibinfo{title}{Diffusion in presence of stochastic resetting at
  power-law times}.
\newblock \emph{\bibinfo{journal}{arXiv preprint arXiv:1512.02092}}
  (\bibinfo{year}{2015}).

\bibitem{pal2015diffusion}
\bibinfo{author}{Pal, A.}
\newblock \bibinfo{title}{Diffusion in a potential landscape with stochastic
  resetting}.
\newblock \emph{\bibinfo{journal}{Physical Review E}}
  \textbf{\bibinfo{volume}{91}}, \bibinfo{pages}{012113}
  (\bibinfo{year}{2015}).

\bibitem{rotbart2015michaelis}
\bibinfo{author}{Rotbart, T.}, \bibinfo{author}{Reuveni, S.} \&
  \bibinfo{author}{Urbakh, M.}
\newblock \bibinfo{title}{Michaelis-Menten reaction scheme as a unified
  approach towards the optimal restart problem}.
\newblock \emph{\bibinfo{journal}{Physical Review E}}
  \textbf{\bibinfo{volume}{92}}, \bibinfo{pages}{060101}
  (\bibinfo{year}{2015}).

\bibitem{reuveni2015optimal}
\bibinfo{author}{Reuveni, S.}
\newblock \bibinfo{title}{Optimal stochastic restart renders fluctuations in
  first passage times universal}.
\newblock \emph{\bibinfo{journal}{arXiv preprint arXiv:1512.01600}}
  (\bibinfo{year}{2015}).

\bibitem{Jahnel:2011uv}
\bibinfo{author}{Jahnel, M.}, \bibinfo{author}{Behrndt, M.},
  \bibinfo{author}{Jannasch, A.}, \bibinfo{author}{Sch{\"a}ffer, E.} \&
  \bibinfo{author}{Grill, S.~W.}
\newblock \bibinfo{title}{{Measuring the complete force field of an optical
  trap.}}
\newblock \emph{\bibinfo{journal}{Optics letters}}
  \textbf{\bibinfo{volume}{36}}, \bibinfo{pages}{1260--1262}
  (\bibinfo{year}{2011}).

\bibitem{Depken:2013dj}
\bibinfo{author}{Depken, M.}, \bibinfo{author}{Parrondo, J. M.~R.} \&
  \bibinfo{author}{Grill, S.~W.}
\newblock \bibinfo{title}{{Intermittent transcription dynamics for the rapid
  production of long transcripts of high fidelity}}.
\newblock \emph{\bibinfo{journal}{{Cell Rep.}}} \textbf{\bibinfo{volume}{5}},
  \bibinfo{pages}{521--530} (\bibinfo{year}{2013}).

\bibitem{gardiner2009}
\bibinfo{author}{Gardiner, C.~W.}
\newblock \emph{\bibinfo{title}{{Stochatic methods}}}
  (\bibinfo{publisher}{Springer-Verlag, Berlin, Germany},
  \bibinfo{year}{2009}).

\bibitem{abramowitz1964handbook}
\bibinfo{author}{Abramowitz, M.} \& \bibinfo{author}{Stegun, I.~A.}
\newblock \emph{\bibinfo{title}{Handbook of mathematical functions: with
  formulas, graphs, and mathematical tables}}.
\newblock \bibinfo{number}{55} (\bibinfo{publisher}{Courier Corporation},
  \bibinfo{year}{1964}).

\bibitem{gillespie1976general}
\bibinfo{author}{Gillespie, D.~T.}
\newblock \bibinfo{title}{A general method for numerically simulating the
  stochastic time evolution of coupled chemical reactions}.
\newblock \emph{\bibinfo{journal}{J. Comp. Phys.}}
  \textbf{\bibinfo{volume}{22}}, \bibinfo{pages}{403--434}
  (\bibinfo{year}{1976}).

\bibitem{redner2001guide}
\bibinfo{author}{Redner, S.}
\newblock \emph{\bibinfo{title}{A guide to first-passage processes}}
  (\bibinfo{publisher}{Cambridge University Press}, \bibinfo{year}{2001}).

\bibitem{Zamft:2012es}
\bibinfo{author}{Zamft, B.}, \bibinfo{author}{Bintu, L.},
  \bibinfo{author}{Ishibashi, T.} \& \bibinfo{author}{Bustamante, C.}
\newblock \bibinfo{title}{{Nascent RNA structure modulates the transcriptional
  dynamics of RNA polymerases}}.
\newblock \emph{\bibinfo{journal}{Proc. Natl. Acad. Sci. U.S.A.}}
  \textbf{\bibinfo{volume}{109}}, \bibinfo{pages}{8948--8953}
  (\bibinfo{year}{2012}).

\bibitem{Abbondanzieri:2005dz}
\bibinfo{author}{Abbondanzieri, E.~A.}, \bibinfo{author}{Greenleaf, W.~J.},
  \bibinfo{author}{Shaevitz, J.~W.}, \bibinfo{author}{Landick, R.} \&
  \bibinfo{author}{Block, S.~M.}
\newblock \bibinfo{title}{{Direct observation of base-pair stepping by RNA
  polymerase}}.
\newblock \emph{\bibinfo{journal}{Nature Cell Biology}}
  \textbf{\bibinfo{volume}{438}}, \bibinfo{pages}{460--465}
  (\bibinfo{year}{2005}).

\end{thebibliography}

\cleardoublepage

\section{Appendix A: Exact solution of the hopping model with diffusion}
\label{app:hoppsol}

Equations~(\ref{eq:me1}-\ref{eq:me3}) can be rewritten as
\begin{equation}
\frac{\text{d}}{\text{d}t} \mathsf{P}(t)  =  \mathsf{A} \mathsf{P}(t)\quad. 
\label{eq:tridiagonal}
\end{equation}
 where  $\mathsf{P}(t) = [p_1(t) \,p_2(t)\,...]^\top$ is a column vector including the state probabilities at time $t$ and $\mathsf{A}$ is a tridiagonal symmetrict Toeplitz matrix~\cite{Golub1996book} of the form
 \[\mathsf{A}= \left[ \begin{array}{ccccc}
-(2k+k_c) & k & 0 & 0 &\hdots \\
 k & -(2k+k_c) & k & 0 & \hdots \ \\
0 & k & -(2k+k_c) & k & 0 \hdots \\
\vdots & \vdots &  \vdots & \ddots & \ddots  \hdots \\
\end{array} \right].\]
The solution of Eq.~\eqref{eq:tridiagonal} with initial condition $\mathsf{P}(0) = [0,0,...,0,1,0,...]^\top$, with $p_{n_0}(0)=1$ and $p_n(0)=0$ for $n \neq n_0$,  is given by \cite{sotomayor1979book}
\begin{equation}\label{eq:solutionGeneral}
\mathsf{P}(t)=\mathsf{P}(0) e^{\mathsf{A}t}\quad.
\end{equation}
We now decompose $\mathsf{A}$ in the following form
\begin{eqnarray}
\mathsf{A}=\mathsf{Q}\,\mathsf{D}\,\mathsf{Q}^{-1}\quad.
\end{eqnarray}
where $\mathsf{D}$ is a diagonal matrix containing the eigenvalues of $\mathsf{A}$ and $\mathsf{Q}$ is a matrix with the eigenvectors of $\mathsf{A}$ in columns. Note that since $\mathsf{A}$ is symmetric, $Q^{-1}=Q^\top$. To obtain the eigenvalues of $\mathsf{A}$ we first assume that the matrix is of finite size $N\times N$ and then take the limit $N\to \infty$. For $N$ finite, the matrix elements of the matrices $\mathsf{D}$ and $\mathsf{Q}$ are given by~\cite{Golub1996book}
\begin{eqnarray}
\mathsf{D}_{ii} & = & -(2k+k_c) + 2k \cos\left(\dfrac{i\pi}{N+1}\right)\quad, \\
\mathsf{Q}_{ij} & = &  \sqrt{\dfrac{2}{N+1}}\sin\left(\dfrac{ij\pi}{N+1}\right)\quad. \label{eq:eigenvector}
\end{eqnarray}
The term $\sqrt{\dfrac{2}{N+1}}$ that appears in Eq. (\ref{eq:eigenvector}) is the normalization constant. As a result, Eq. \eqref{eq:solutionGeneral} can be rewritten as
 \begin{equation}\label{eq:solutionG}
\mathsf{P}(t)= \mathsf{Q}\,e^{\mathsf{D}t}\,\mathsf{Q}^{\top} \, \mathsf{P}(0)\quad. 
\end{equation}
where $e^{\mathsf{D}t}$ is a diagonal matrix with elements $e^{\mathsf{D}_{ii} t}$ ($i=1,\dots,N$) in the diagonal. After some  algebra, we obtain the following expression for the $n-$th element of the vector $\mathsf{P}(t)$
 \begin{eqnarray}
 p_n (t) & = & \displaystyle \sum_{m=1}^N \sin\left(\dfrac{kn\pi}{N+1}\right) e^{\left[ -(2k+k_c) + 2k \cos\left(\frac{m\pi}{N+1}\right) \right]t}\times \nonumber \\
 &\times& \dfrac{2}{N+1}\sin\left(\dfrac{n_0k\pi}{N+1}\right)\quad.
 \end{eqnarray}
 
 We now take the asymptotic limit $N\to \infty$. In this limit, $k/N\to x$ where $x$ is a continuous variable and the sum $\sum_{k=1}^N \frac{1}{N}\to \int_0^1 \text{d} x$. Using these approximations, we obtain
 \begin{eqnarray}
  p_n(t) & = &  \frac{2 e^{-(2k+k_c)t}}{\pi} \int_0^\pi  \, e^{ 2kt\cos{ x}} \sin(n_0x) \sin(nx)\,\text{d}x \nonumber \\
  & = & \frac{e^{-(2k+k_c)t}}{\pi} \left\{\int_0^\pi  e^{ 2kt\cos{ \left( x\right)}} \cos[(n_0-n)x]\,\text{d}x \right.\nonumber\\
 &-&  \int_0^\pi   \left. e^{ 2kt\cos{ \left( x\right)}}\cos[(n_0+n)x]  \, \text{d}x\right\} \nonumber \\
 &=&  e^{-(2k+k_c)t}[I_{n_0-n}(2kt)-I_{n_0+n}(2kt)]\quad. \label{eq:pj}
 \end{eqnarray} 
 Here we have used the property
  \begin{eqnarray}
 \sin(ax)\sin(bx) &=& \dfrac{1}{2} \left\{\cos[(a-b)x]-\cos[(a+b)x]\right\},\nonumber
 \end{eqnarray}
and the definition of the  modified Bessel function of the first kind~\cite{abramowitz1964handbook}
 \begin{equation}
 I_m(z) =  \displaystyle \dfrac{1}{\pi} \int_{0}^{\pi}   e^{z \cos x} \cos(mx)\,\text{d}x \quad.
 \end{equation}
 From Eq.~\eqref{eq:pj} we can obtain the probability to be at state $n=1$ at time $t$,
\begin{equation}
p_1(t) =  e^{-(2k+k_c)t}\,[I_{n_0-1}(2kt)-I_{n_0+1}(2kt)]\quad. \label{eq:p11}
\end{equation}
Using the property $I_{m-1}(z)-I_{m+1}(z) = \frac{2m}{z}I_m(z)$~\cite{abramowitz1964handbook} in Eq.~\eqref{eq:p11} we obtain
\begin{equation}
p_1(t) =  e^{-(2k+k_c)t}\frac{n_0 I_{n_0} (2kt)}{kt}\quad, \label{eq:p12}
\end{equation}
which equals to Eq.~\eqref{eq:p0}.

 \section{Appendix B: Calculation of the mean recovery time from the Backward Fokker-Planck equation}
The Backward Fokker-Planck equation corresponding to Eq.~\eqref{eq:FPE} in the continuous-space model reads
\begin{equation}
\frac{\partial\rho(x,t|x_0,0)}{\partial t} = D\frac{\partial^2\rho(x,t|x_0,0)}{\partial x_0^2} - k_c \rho(x,t|x_0,0)\quad.
\label{eq:BFPE}
\end{equation}
Integrating Eq.~\eqref{eq:BFPE} with respect to $x$ from $x=0$ to $x=\infty$ we obtain the following equation for the survival probability:
\begin{equation}
\frac{\partial S(t;x_0)}{\partial t} = D \frac{\partial^2 S(t;x_0)}{\partial x_0^2} - k_cS(t;x_0)\quad.
\label{eq:BFPE}
\end{equation}
Taking the Laplace transform, Eq.~\eqref{eq:BFPE} yields
\begin{equation}
q\mathbb{S}(q;x_0)-1 = D \frac{\partial^2 \mathbb{S}(q;x_0)}{\partial x_0^2} - k_c\mathbb{S}(q;x_0)\quad,
\label{eq:LBFPE}
\end{equation}
where $\mathbb{S}(q;x_0) = \int_0^{\infty} \text{d}t e^{-qt} S(t;x_0)$ is the Laplace transform of the survival probability and we have used $S(0;x_0)=1$. The solution of Eq.~\eqref{eq:LBFPE} is given by
\begin{equation}
\mathbb{S}(q;x_0) =\frac{1}{k_c+q}\left[1-e^{-x_0\sqrt{(k_c+q)/D}}\right]\quad.
\label{eq:LS}
\end{equation}
From Eq.~\eqref{eq:LS} one can find all the moments of the recovery time distribution. In particular, the mean recovery time:
\begin{equation}
\langle\tau_{\rm rec}\rangle=\mathbb{S}_n(0) =\frac{1}{k_c}\left[1-e^{-x_0\sqrt{k_c/D}}\right]\quad,
\label{eq:LS2}
\end{equation}
which coincides with Eq.~\eqref{eq:meanRTSI}.\cb

\end{document}